\documentclass[aps,prf,twocolumn,superscriptaddress,9pt,usenames,dvipsnames]{revtex4-1}
\usepackage{amsmath}
\usepackage{graphicx}
\usepackage{mathtools}
\usepackage{mathtools}
\usepackage{stmaryrd}
\usepackage{float}
\usepackage{empheq}
\usepackage{epstopdf}
\usepackage{bigints}
\usepackage{epstopdf}
\usepackage{amssymb}
\usepackage{wrapfig}
\usepackage{amsmath,hyperref}
\usepackage[noabbrev]{cleveref}
\usepackage{soul}
\usepackage{textcomp}
\usepackage{mydef}
\usepackage{xcolor}
\usepackage{comment}
\usepackage[makeroom]{cancel}
\usepackage{bm}

\hypersetup{
colorlinks,
citecolor=blue,
filecolor=red,
linkcolor=blue,
urlcolor=blue,
hyperfigures
}

\providecommand \selectlanguage [0]{\@gobble}%
\providecommand \bibinfo  [0]{\@secondoftwo}%
\providecommand \bibfield  [0]{\@secondoftwo}%
\providecommand \BibitemShut  [1]{\csname bibitem#1\endcsname}%

\begin{document}

\title{Emergent odd response in active chiral films}

\author{Seema Chahal}
\thanks{These two authors contributed equally to this work.}
\affiliation{International Centre for Theoretical Sciences, Tata Institute of Fundamental Research,
Shivakote, Bengaluru 560089, India}

\author{D. Naveen Kumar}
\thanks{These two authors contributed equally to this work.}
\affiliation{International Centre for Theoretical Sciences, Tata Institute of Fundamental Research,
Shivakote, Bengaluru 560089, India}

\author{Brato Chakrabarti}
\email{brato.chakrabarti@icts.res.in} 
\affiliation{International Centre for Theoretical Sciences, Tata Institute of Fundamental Research,
Shivakote, Bengaluru 560089, India}

\
\date{\today}

\begin{abstract}
Active chiral fluids can support an anomalous transport coefficient known as odd (or Hall) viscosity. Hydrodynamic descriptions of such fluids typically introduce odd viscosity phenomenologically. How such a response emerges from specific microscopic interactions remains incompletely understood. Here, building on classical shear rheology, we microscopically derive an odd rheological response in active chiral films: thin layers of torque-exerting, elongated particles anchored to a no-slip surface. A canonical realization of such a film is the bacterial carpet, in which flagellated bacteria are tethered head-down to a solid surface while their flagella remain free to spin and inject angular momentum into the surrounding fluid. Using a Smoluchowski kinetic theory for the orientational dynamics of these anchored particles, we derive their stress response to an imposed shear flow. We reveal that shear-induced reorientation leads to a flow-aligned polarization and a transverse surface traction from which the odd-viscosity tensor follows in closed form. This approach stands in contrast with existing microscopic theories, in which the axis of torque injection is insensitive to the flow rather than set by it. Numerical solutions of the kinetic theory further highlight saturation of the transverse traction at strong shear, driven by this orientation dynamics -- signaling departure from linear response. Our results demonstrate how odd viscosity can emerge self-consistently as a coarse-grained rheological signature of active fluid-structure interaction and establish active chiral films as a new controllable setting for odd hydrodynamics.
\end{abstract}
\pacs{...}
\maketitle

\vspace*{-5mm}
\section{Introduction}\label{sec:Introduction}
Viscosity is conventionally understood as a dissipative response: an applied strain rate generates a stress that opposes relative motion between fluid layers, converting mechanical energy to heat. Formally, the viscosity tensor $\eta_{ijkl}$ is the linear-response coefficient relating the viscous stress $\sigma_{ij}^{\rm vis}$ to the local rate-of-strain $E_{kl}$ as $\sigma_{ij}^{\rm vis} = \eta_{ijkl}E_{kl}$. Time-reversal symmetry {in ordinary fluids}, via Onsager reciprocity, requires this tensor to obey the major (block) symmetry under the permutation $\{ij\} \leftrightarrow \{kl\}$, so that the resulting viscous dissipation, ${\Phi = } \sigma_{ij}^{\rm vis} E_{ij}$, is governed by a positive semi-definite quadratic form \citep{de2013non}. {In chiral fluids, which are often characterised by a time-reversal-odd quantity such as an intrinsic angular-momentum density $\bm{\ell}$, microscopic reversibility implies the Onsager--Casimir relation $\eta_{ijkl}(\bm{\ell}) = \eta_{klij}(-\bm{\ell})$ \citep{casimir1945onsager, everts2024dissipative}. At fixed $\bm{\ell}$ the major symmetry is then no longer enforced, and the viscosity tensor may acquire a major-antisymmetric part $\eta^o_{ijkl} = (\eta_{ijkl} - \eta_{klij})/2$, known as the odd (or Hall) viscosity. This contribution generates a stress transverse to the applied shear and does not enter the viscous dissipation directly, since $\Phi = \eta_{ijkl}E_{kl}E_{ij}$ contracts the viscosity tensor with an object symmetric under $\{ij\} \leftrightarrow \{kl\}$, so that only the major-symmetric part survives \citep{fruchart2023odd, khain2022stokes}. Interestingly, because odd viscosity  alters the flow field itself, in three dimensions it can increase the total dissipation relative to an equivalent system without it, for the same boundary conditions \citep{everts2024dissipative, meissner2026exact}.} First identified in the context of quantum Hall fluids \citep{PhysRevLett.75.697}, odd viscosity has since been recognized as a much more general hydrodynamic phenomenon with examples including polyatomic gases in a magnetic field \citep{beenakker1970magnetic}, chiral superfluids and superconductors \citep{read2011hall}, two-dimensional vortex fluids \citep{wiegmann2014anomalous}, and — most relevant to the present work — chiral active matter, in which the microscopic constituents themselves have spin degrees of freedom \citep{furthauer2012active, petroff2015fast, banerjee2017odd, soni2019odd, tan2022odd, hosaka2023hydrodynamics, markovich2024nonreciprocity}. Recent work has further revealed the unique consequences of odd viscosity for both passive and active particle dynamics. Examples range from counterintuitive trajectories of sedimenting spheres \citep{khain2022stokes}, modified suspension rheology in odd fluids \citep{everts2024dissipative, yosaka2026odd}, and propulsion of microswimmers with torque dipoles \citep{hosaka2023lorentz}, to asymmetric spreading and chiral flows in highly deformable odd droplets \citep{francca2026odd}.

A central and comparatively underexplored challenge in the study of odd viscous fluids is connecting the odd viscosity tensor itself to the microscopic dynamics of the constituents that generate it — whether spinning magnetic colloids \citep{soni2019odd}, chiral bacterial crystals \citep{petroff2015fast}, or torqued active dumbbells \citep{markovich2024nonreciprocity}. Many studies instead treat the description of such a chiral fluid phenomenologically \citep{ganeshan2017odd} and explore its consequences, ranging from anomalous edge and chiral waves \citep{soni2019odd, poggioli2023emergent} to odd turbulence \citep{de2024pattern}.  While it is well appreciated that breaking time-reversal and parity symmetries permits such a coefficient to exist in chiral fluids, far less is known about how it manifests at the macroscale from specific microscopic interactions. Recent work has begun to close this gap from several directions. Approaches include using Onsager's regression hypothesis to relate odd viscosity to the time-antisymmetric stress fluctuations in active fluids \citep{epstein2020time}; coarse-graining kinetic energy of rigid bodies with sustained spin angular momentum density \citep{markovich2024nonreciprocity}; and applying Boltzmann kinetic theory to granular gas with parity-violating, chiral collisions between microscopic constituents \citep{eren2025collisional}.

\begin{figure*}[htb]
    \centering
    \includegraphics[width=1\linewidth]{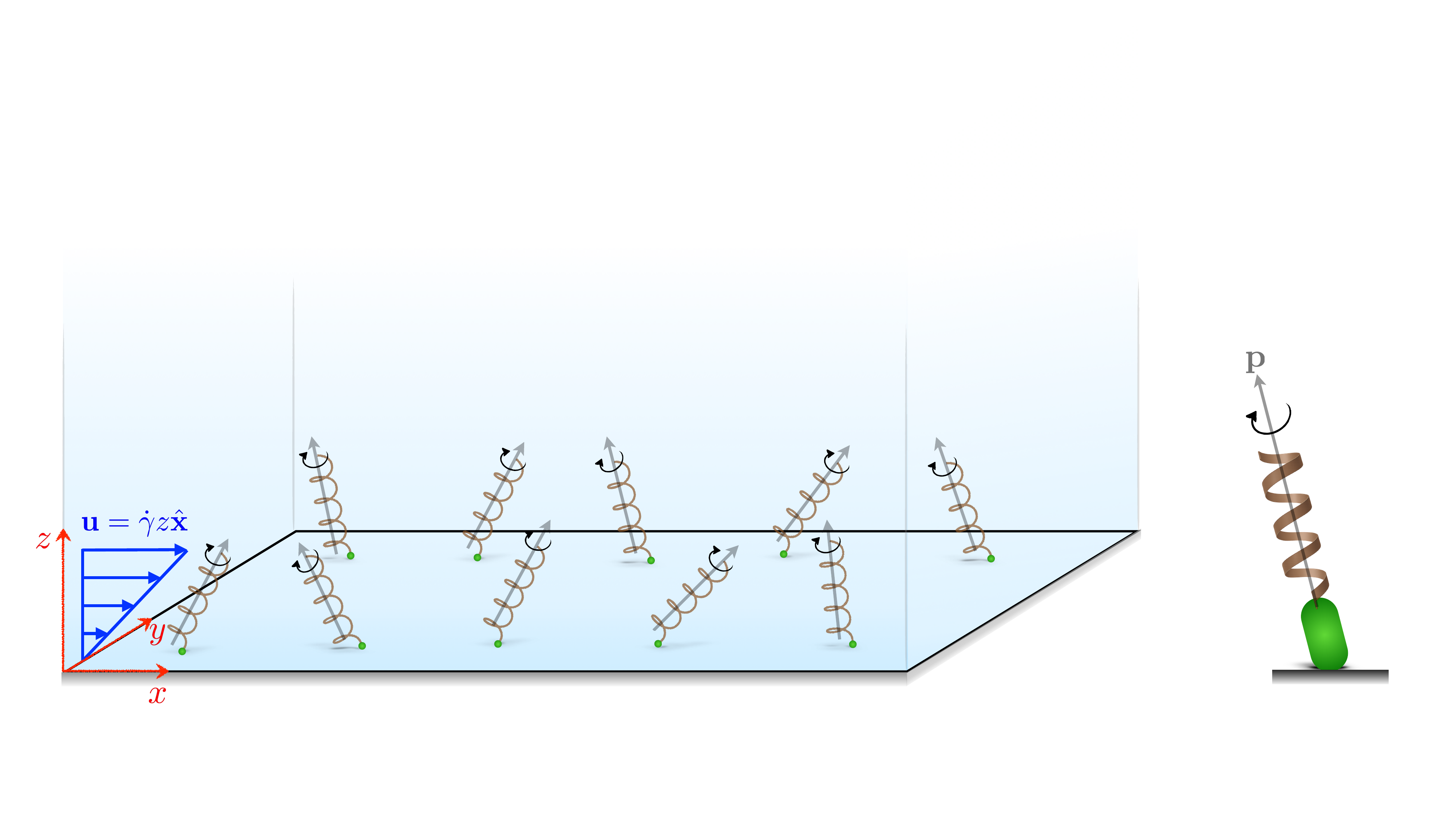}
    \caption{Schematic of a bacterial carpet -- a canonical active chiral film. Flagellated bacteria are anchored to a no-slip surface at $z=0$. Each flagellum spins continuously, injecting angular momentum into the surrounding fluid. The orientation of the flagellum (on the right) is described by a unit vector $\bm{p}(\theta,\phi)$, expressed in the local spherical basis $\{\uvc{e}_\theta,\uvc{e}_\phi\}$, which acts as its spin axis and is constrained to lie in the upper hemisphere $\mathcal{H}$. To probe the rheological response of the film, an external shear flow is imposed to reorient the particles.}
    \label{fig:fig1_schematic}
\end{figure*}

In this work, we propose a new approach that builds on a Smoluchowski kinetic theory framework of shear rheology \citep{Doi1986Theory} to probe the microscopic origin of odd response in active chiral films \citep{furthauer2012active}. Such an active film is composed of elongated, torque-exerting particles anchored to a surface, powered either by internal biological machinery~\citep{Berg2003RotaryMotor} or external actuation, and immersed in a Stokesian fluid of viscosity $\mu$. Anchoring suppresses lateral migration, leaving the particle orientation as the sole dynamical degree of freedom — a configuration often termed an \textit{active carpet}~\citep{PhysRevLett.121.248101, chakrabarti2024pnas, naveen2026growing}. A canonical and experimentally accessible realization of such active carpets is the bacterial carpet~\citep{Darnton2004BacterialCarpets}, in which flagellated bacteria are tethered head-down to a solid surface (see figure~\ref{fig:fig1_schematic}). The flagellar bundle remains free to spin, continuously injecting angular momentum into the surrounding  fluid~\citep{Dahler1961AngularMomentum} with its hydrodynamic signature being that of a \textit{rotlet} in Stokes flow. Each bacterium thus acts as a torque monopole whose axis is set by the instantaneous orientation of the flagellum, described by a unit vector $\bm p$. As in classical shear rheology, we probe the response of this active film to an imposed simple shear flow (see figure~\ref{fig:fig1_schematic}). As the flagellum reorients with the ambient flow, the axis of torque injection reorients with it, making orientation the degree of freedom that governs the rheological response. Thus our approach differs from other descriptions of odd viscous flows, in which the injected torque density --- from which the odd response inherits its axis --- is insensitive to the macroscopic flow. Instead, here, we reveal that the shear-induced reorientation dynamics of the active particles is central to giving rise to an emergent Hall-like response, allowing us to derive the full viscosity tensor microscopically without appealing to phenomenology.

The remainder of this paper is organized as follows. In Sec.~\ref{sec:kintheory}, we outline the kinetic theory of the active film and derive its volume-averaged particle stress. Section~\ref{sec:rheology} then obtains the emergent viscosity tensor as a linear-response coefficient under weak shear and elucidates connections with the classical rheology of passive rod suspensions. We subsequently examine the {nonlinear} regime at high shear rates, where simulations reveal a saturating Hall traction that departs from the typical linear response. We conclude in Sec.~\ref{sec:conclusion} by discussing experimental implications and directions for future work.

\vspace*{-2mm}
\section{Kinetic theory for active chiral film}\label{sec:kintheory}

The active chiral film considered here is assumed to be thin in the long-wavelength sense: the characteristic height $h$ of the active layer is set geometrically by the contour length of the anchored particles, $h \sim \ell$ (with $\ell \simeq 8$–$10 \ \mu$m for typical \emph{E.~coli} flagella), and is much smaller than its lateral extent $L$. Thus, the in-plane collective response captures the dominant physics, and the hydrodynamics is well-described in the half-space $z>0$. To probe the rheology of this film, we adopt a kinetic theory framework \citep{Doi1986Theory}, which allows us to derive the emergent rheological response directly from the microscopic orientational dynamics of the particles. In this description, the particle configuration within the film is encoded in a distribution function $\psi(\bx_\pll,\bp,t)$, where $\bx_\pll = \{x,y\}$ is the surface coordinate along the no-slip surface at $z = 0$, and $\bp$ is the unit vector characterizing the orientation of the particles (see figure~\ref{fig:fig1_schematic}). 
Throughout, Cartesian indices $\{1,2,3\}$ correspond to $\{x,y,z\}$, and we denote the unit wall normal by $\uvc{n} \equiv \uvc{z}$, with the associated surface projector $\bm{\Pi} = \bm{I} - \uvc{n}\,\uvc{n}$.
The distribution function encodes the probability density of finding a particle at the anchoring position $\bx_\pll$ on the surface with orientation $\bp$ at time $t$. Importantly, since the particles are anchored to and cannot penetrate the no-slip surface, their orientations are strictly confined to the upper hemisphere
\begin{equation}\label{eq:hemisphere}
    \mathcal{H} \equiv \mathbb{S}^2_+ = \{\bp \in \mathbb{S}^2 : \bp \cdot \hat{\bm{z}} \ge  0\},
\end{equation}
consistent with the half-space nature of the hydrodynamics. This constraint, while geometrically natural, has nontrivial consequences for the orientational dynamics and will play a central role in determining the rheology of the active film. To probe the response of the thin film, we now impose an external simple shear flow $\bu = \dot{\gamma} z \hat{\bx}$, where $\dot{\gamma}$ is the shear rate.
This choice is not restrictive. Consider an incompressible linear flow near the anchoring surface with $u_i = A_{ij}x_j$, where $\bm{A} = \nabla\bm{u}$ is the constant velocity-gradient tensor. If the flow satisfies the no-slip and impermeability conditions on the wall, it is necessarily a simple shear. No-slip eliminates all in-plane gradients, $\bm{\Pi}\cdot\bm{A}\cdot\bm{\Pi} = \bm{0}$ and $\uvc{n}\cdot\bm{A}\cdot\bm{\Pi} = \bm{0}$; incompressibility then forces $\uvc{n}\cdot\bm{A}\cdot\uvc{n}=0$, leaving $\bm{A} = \dot{\boldsymbol{\gamma}}\otimes\uvc{n}$, with the in-plane shear-rate vector $\dot{\boldsymbol{\gamma}}$ ($\dot{\boldsymbol{\gamma}}\cdot\uvc{n}=0$) as the sole surviving kinematic content. Since our setup possesses continuous rotational symmetry about $\uvc{n}$, we may align $\dot{\boldsymbol{\gamma}} = \dot{\gamma}\,\uvc{x}$ without any loss of generality.
In addition to this imposed flow, each anchored particle also generates a disturbance flow in the surrounding fluid via its rotlet singularity. However, in the dilute limit, the concentration $c \ll 1$ and the disturbance flow generated by the active film is $O(c)$ relative to the $O(1)$ imposed external shear. Thus, we neglect hydrodynamic interactions between the particles and account for only the ambient shear flow $\bu$ that drives the orientational dynamics within the film. This approximation is similar to that used in prior work on active suspensions, where the rheological response is computed from the single-particle orientational dynamics in a background flow~\citep{PhysRevLett.92.118101, saintillan2018rheology}. A further consequence of this dilute limit is that the only flow experienced by each particle is the spatially uniform imposed shear, which carries no information about the particle's anchoring position $\bm{x}_\parallel$ and is unable to generate spatial gradients in $\psi$. The distribution function thus reduces to $\psi(\bx_\pll,\bm{p},t)\equiv\psi(\bm{p},t)$, with the following normalization condition
\begin{equation}\label{eq:normalsation_psi}
\int_{\mathcal H} \psi(\bm p,t)\,\mathrm d\bm p = 1.
\end{equation}
The dimensionless evolution equation of $\psi$ satisfies a conservation or a Smoluchowski equation \citep{Doi1986Theory}
\begin{equation}\label{eq:psi_conservation}
\partial_t\psi+\nabla_{\bm p}\cdot\bm J_{\bm p}=0,
\end{equation}
where $\nabla_\bp = (\bm{I} - \bp \bp) \cdot \partial_\bp$ is the surface gradient on the unit sphere and $\bm{J}_{\bm{p}}$ is the orientational flux, given by
\begin{equation}\label{eq:rotational_flux_Jp}
\bm J_{\bm p}= Pe \,\dot{\bm p}\psi-\nabla_{\bm p}\psi.
\end{equation}
In the above expression, the first term represents the advective flux due to shear-induced reorientation, while the second term accounts for Brownian rotational diffusion that acts to isotropize the orientation distribution. The relative importance of these competing effects is captured by a Péclet number defined as 
\begin{equation}
Pe = \frac{\dot{\gamma}}{d_r},
\end{equation}
where $d_r$ is the rotational diffusivity of a single particle; as evident, here, we have used the diffusive timescale $d_r^{-1}$ to nondimensionalize the governing equations. The reorientation of each particle in the imposed flow is governed by Jeffery's equation~\citep{jeffery1922motion}, which describes the rotation of a slender axisymmetric body in a viscous flow
\begin{equation}\label{eq:pdot_Jeffery}
\dot{\bm p}=(\bm I-\bm p\bm p) \cdot \left(\lambda \bm{E} + \bm{W}\right)|_{z=0} \cdot \bm p.
\end{equation}
Here $\{\bm{E},\bm{W}\}$ are respectively the rate-of-strain and rate-of-vorticity tensors of the shear flow, and $\lambda$ is the Bretherton constant characterizing the shape anisotropy of the particle. While Jeffery's equation was originally derived for a free particle in a shear flow, an analogous dynamical equation appropriate for the present problem has recently been derived for pinned rod-shaped particles \citep{chakrabarti2024pnas}. Furthermore, for an axisymmetric body, the rotational resistance tensor is diagonal in the frame of $\bm{p}$, so that spin about the symmetry axis decouples from tilting of that axis. Although a helical flagellum possesses a rotation--translation coupling, the resulting thrust is directed along $\bm{p}$ and exerts no couple about the anchoring point; its resistance tensor is also transversely isotropic about the helix axis once averaged over a turn. Thus, we highlight that the active axial torque in either situation drives only spin about $\bm{p}$ without affecting its reorientation dynamics. For the sake of simplicity, we will adopt $\lambda = 1$, which corresponds to slender rod-like particles as appropriate for the flagellum in figure~\ref{fig:fig1_schematic}. This is also the regime in which the constraint \eqref{eq:hemisphere} is sharpest, since $\mathcal{H}$ follows from the wall sterically excluding the particle body, which extends along $\bm{p}$. We highlight that, in our framework, one can also account for an additional restoring torque in \eqref{eq:pdot_Jeffery} akin to a torsional spring arising from the pinning constraint itself: an effect we do not incorporate here. Such a passive contribution only quantitatively alters the shape of the steady-state distribution and does not qualitatively change our conclusions in what follows. Equations~\eqref{eq:psi_conservation}--\eqref{eq:pdot_Jeffery}, together with appropriate boundary conditions on $\mathcal{H}$, fully specify the orientational dynamics of the particles. Parameterizing the orientation vector as $\bm{p} = (\cos\phi\sin\theta, \sin\phi\sin\theta, \cos\theta)$ in terms of the polar angle $\theta \in [0, \pi/2]$ and azimuthal angle $\phi \in [0, 2\pi)$, the distribution function  $\psi(\theta, \phi, t)$ can be seen to be periodic in the azimuthal direction,
\begin{equation}
\psi(\theta, \phi) = \psi(\theta, \phi + 2\pi).
\end{equation}
At the equator $\theta = \pi/2$, which corresponds to $z=0$, the boundary condition is set by the physical requirement that particles cannot penetrate the wall. At $\theta = \pi/2$, the polar unit vector $\hat{\bm{e}}_\theta|_{\theta=\pi/2} = -\hat{\bm{z}}$ points into the lower hemisphere (see figure~\ref{fig:fig1_schematic}), so the requirement of no probability flux through the wall translates to,
\begin{equation}\label{eq:no_flux_bc}
\bm{J}_{\bm{p}} \cdot \hat{\bm{z}}\big|_{\theta = \pi/2} = 0.
\end{equation}
This boundary condition is the natural choice consistent with the hemisphere constraint $\mathcal{H}$ introduced earlier, and together with the periodicity in $\phi$ and the appropriate normalization condition defined in \eqref{eq:normalsation_psi}, ensures that $\psi$ remains a well-defined probability density on $\mathcal{H}$ for all time. 

Having established the orientational dynamics of the carpet, we now turn to its rheological response. The macroscopic particle stress generated within the thin film is obtained by volume averaging~\citep{Batchelor_1970}, which relates the bulk rheological properties of the film to the single-particle mechanics. In our framework, the only channel of activity is the torque monopole exerted by each spinning particle on the fluid. The origin of the particle stress is therefore the body-torque density that the flagellar motors impose on the fluid, and the balance of angular momentum fixes the stress in terms of it, giving rise to a purely antisymmetric particle stress~\citep{condiff1964fluid, seema2026}. We note that angular momentum is not conserved within the film itself, since the flagella are anchored and the reaction torque is transmitted to the substrate --- this is precisely why the active singularity here is an unpaired torque monopole rather than the torque dipole as seen in bulk active fluids. The dimensional volume-averaged particle stress is thus obtained as $\Sigma^{\ast}_{ij}=\frac{\bar n}{2}\epsilon_{ijk}\langle T_k(\bp)\rangle$, where $\bar n$ is the surface density of active particles, $\epsilon_{ijk}$ is the Levi-Civita tensor, and $\bm{T}(\bp)=\beta\bp$ is the torque per unit body length exerted by an individual particle on the fluid, with $\beta$ being its characteristic magnitude and assumed to be identical for all the particles. Here, $\langle\cdot\rangle=\int_{\mathcal H}(\cdot)\,\psi(\bp)\,\mathrm{d}\bp$ denotes an orientational average. Importantly, while the film is quasi-two-dimensional in aspect ratio, the stresses within it remain genuinely three-dimensional, volume-averaged quantities. We measure stresses in units of the viscous stress scale $\mu d_r$ and define a dimensionless activity $\mathcal{A}\equiv\bar n\beta/(\mu d_r)$. To make explicit that $\mathcal{A}$ carries information about microscopic content, we estimate it for a carpet of anchored \textit{E.\ coli} bacteria in an aqueous environment. For a given spin rate of an isolated rotating flagellum, a resistive-force-theory estimate of the axial torque density provides $\beta\simeq4\times10^{-14}\,\mathrm{N}$ and $d_r\simeq0.04\,\mathrm{rad^2/s}$ \citep{chattopadhyay2006swimming}; combined with reported carpet densities  $\bar n\simeq0.05$--$0.2\,\mu\mathrm{m}^{-2}$ \citep{Darnton2004BacterialCarpets}, we obtain $\mathcal{A}=\bar n\beta/(\mu d_r)\sim O(50\text{--}100)$. As $\beta$ and $d_r$ are computed microscopically, and $\bar n$ is estimated from experiments similar to the volume fraction in the Einstein viscosity law, no phenomenological parameter enters $\mathcal A$. Henceforth, $\Sigma_{ij}\equiv\Sigma^{\ast}_{ij}/(\mu d_r)$ denotes the dimensionless particle stress, which takes the form
\begin{equation}\label{eq:active_stress}
\Sigma_{ij}=\frac{\mathcal A}{2}\epsilon_{ijk}P_k,
\end{equation}
where $\bP=\langle\bp\rangle$ is the polarity or the magnetization of particles in the active film. In analogy with the rheology of passive rod-suspensions~\citep{leal1971effect}, the sections that follow illustrate how the orientational dynamics and the associated particle stress lead to an emergent viscosity tensor for the active film, carrying clear signatures of odd response. We note that while here we focus on the active particle stresses, the pinning of the particles onto the surface also produces a constraint force-monopole, reacting against the hydrodynamic drag on the anchored particle. Similarly, at the next order in hydrodynamic signatures, the particles produce a constraint stress in the thin film, akin to classical suspensions of rigid rods~\citep{hinch1972effect}. These passive contributions do play a role in the entire rheological response of the thin film; however, they leave the odd signatures identified below unaltered.

\vspace*{-2mm}
\section{Emergent odd response}\label{sec:rheology}

\subsection{Weak $Pe$ asymptotics and linear response}

We begin by characterizing the weak-shear response of the active film for $Pe \ll1$. Seeking a steady orientational distribution, we expand perturbatively
\begin{equation}\label{eq:psi_expansion}
\psi=\psi^{(0)}+Pe\,\psi^{(1)}+Pe^{2}\,\psi^{(2)}+\cdots.
\end{equation}
At leading order, rotational diffusion dominates and yields the isotropic distribution $\psi^{(0)}=1/2\pi$. The associated particle stress from \eqref{eq:active_stress} is then obtained as $\Sigma_{ij}^{(0)}= \frac{\mathcal{A}}{4} \epsilon_{ij3}$, where the particle stress has been expanded analogously to \eqref{eq:psi_expansion}. Although nonzero, this leading-order stress is spatially uniform and exerts no traction on the no-slip surface, since $\Sigma_{ij}^{(0)}\hat{z}_j=\Sigma_{i3}^{(0)}=0$. It therefore generates no disturbance flow and no measurable rheological response. The first nontrivial contribution consequently arises at $O(Pe)$ through the correction $\psi^{(1)}$, which we now compute.

On using \eqref{eq:psi_conservation} and collecting terms at $O({Pe})$ in the steady state, we find that $\psi^{(1)}$ satisfies
\begin{equation}
\label{eq:psi1_noflux_eqn}
\Delta_\bp \psi^{(1)}=-\frac{3}{2\pi}\,\bm{pp}:\bm E,
\end{equation}
where $\Delta_\bp$ is the Laplace-Beltrami operator on the unit sphere. This equation is supplemented by the no-flux condition $\partial_\theta\psi^{(1)}|_{\theta=\pi/2}=0$ at the equator and by regularity at the pole $\theta=0$. To solve for $\psi^{(1)}$, we exploit the periodicity in the azimuthal direction of the forcing on the right-hand side of \eqref{eq:psi1_noflux_eqn} and expand the perturbation in a Fourier basis as
\begin{equation}
\label{eq:fourier_ansatz}
\psi^{(1)}(\theta,\phi)=\sum_{m=-\infty}^{\infty}f_m(\theta)e^{im\phi},
\end{equation}
where $m \in \mathbb{Z}$. Using this Ansatz in \eqref{eq:psi1_noflux_eqn}, one finds that only the $m=\pm1$ modes are excited by the imposed shear, each satisfying the same second-order ordinary differential equation for {$f_{\pm 1}(\theta)$}. The resulting distribution, accurate to $O(Pe)$, is then obtained as
\begin{equation}
\label{eq:psi_noflux_solution}
\psi=\frac{1}{2\pi}+\frac{Pe}{4\pi}\frac{\cos\phi}{\sin\theta}\left(1-\cos^3\theta\right).
\end{equation}
The corresponding polarity $\bm P=\langle\bm p\rangle$ follows directly from \eqref{eq:psi_noflux_solution} as
\begin{equation}\label{eq:polarity}
\bm P=\frac{1}{2}\hat{\bm z}+\frac{3}{16}\,Pe\,\hat{\bm x},
\end{equation}
which, through \eqref{eq:active_stress}, provides us with the dimensionless particle stress. We now demonstrate how this polarization encodes the odd response of the active film. The first term in \eqref{eq:polarity} represents a spontaneous polarization normal to the surface: restricting orientations to the upper hemisphere $\mathcal H$ breaks the up-down symmetry and biases the mean orientation toward $\hat{\bm z}$ even in the absence of flow. The second term is the shear-induced polarization along the flow direction $\hat{\bm x}$; it grows linearly with $Pe$ and is intrinsically tied to the origin of the odd-viscous stress in the carpet. To make this explicit, we compute an experimentally tractable quantity: the traction exerted on the no-slip surface in the direction transverse to the imposed shear flow.  In a standard viscous fluid under simple shear, as shown in figure~\ref{fig:fig1_schematic}, the only nonzero traction on the wall lies in the plane of shear, $\Sigma_{xz} \neq 0$, while the transverse component vanishes identically. Here, however, we find 
\begin{equation}\label{eq:ty_linear}
    \Sigma_{yz} = t_y = \frac{3 \mathcal{A}}{32}\,Pe \equiv \frac{\mathcal{A}}{2}\bm P\cdot\hat{\bm x}.
\end{equation}
This transverse traction, which results from shear-induced polarization along the flow direction, is precisely the mechanical counterpart of the Hall voltage in charge transport, where a longitudinal current induces a transverse voltage --- an analogy well recognized in quantum Hall fluids~\citep{PhysRevLett.75.697}. Evidently, the transverse traction increases linearly with the dimensionless activity $\mathcal{A}$, which quantifies the strength of injection of angular momentum in the thin film and also changes sign upon reversing the microscopic sense of rotation of the constituents. The above result further admits a natural interpretation in terms of linear response, wherein $t_y$ grows linearly with $Pe$ in the weak-shear regime, and one identifies 
\begin{equation}
    \eta^o = \frac{t_y}{Pe} = \frac{3 \mathcal{A}}{32},
\end{equation}
as the measure of the odd-viscosity coefficient, or equivalently, the Hall transport coefficient. Taken together, our results highlight how flow-induced reorientation of spinning elongated particles gives rise to rheological signatures of Hall viscosity in an active chiral film. While our results are most readily realized for prolate particles with $\lambda > 0$, we believe our mechanism to be more general. For oblate particles with $\lambda \in [-1,0)$, the symmetry axis $\bp$, now the normal to the particle, still tilts into the flow direction, and an analogous polarization along $\uvc{x}$ should follow. Its magnitude, however, depends on the admissible orientational domain set by steric exclusion, which is no longer the hemisphere $\mathcal{H}$.

While the transverse traction provides a scalar diagnostic of the odd response, the same linear-response framework allows us to reconstruct the full viscosity tensor of the dilute active film. To this end, we first note that the linear dependence of $\bm{P}$ on $Pe$ in the weak-shear regime suggests a general relation between the flow-induced polarity and the applied strain rate, written in terms of the wall normal $\uvc{n}$ ($\equiv \uvc{z}$ here) as
\begin{equation}
	P_k = \frac{1}{2}\, n_k + \frac{3 Pe}{8}\, E_{kl}\, n_l.
	\label{eq:polgen}
\end{equation}
The structure of \eqref{eq:polgen}, the polarization responding to the rate of strain alone, is inherited from the kinetic theory rather than assumed. In the $O(Pe)$ problem of \eqref{eq:psi1_noflux_eqn}, the velocity gradient forces $\psi^{(1)}$ only through the symmetric contraction $\bp\bp:\bm{E}$. The vorticity merely co-rotates orientations rigidly, and it sources no linear response about the isotropic state; the no-flux condition at the equator also remains homogeneous for any flow admissible above the no-slip wall. Linearity then yields $\psi^{(1)} = \chi_{kl}(\bp)\, E_{kl}$ with a linear response kernel $\chi_{kl}(\bp)$ symmetric in $\{k,l\}$ by construction, whose first orientational moment is the $O(Pe)$ contribution in \eqref{eq:polarity}. Substituting the general form of polarity in \eqref{eq:active_stress}, we find the coefficient of the first non-trivial stress contribution at $O(Pe)$ as

\begin{equation}
	\Sigma^{(1)}_{ij} = \left(\frac{3\mathcal{A}}{16}\, \epsilon_{ijk}\, n_l\right) E_{kl}.
\end{equation}
The above expression allows us to infer the viscosity tensor of the active film as the linear-response coefficient relating $\Sigma^{(1)}_{ij}$ and $E_{kl}$; symmetrizing over the strain-rate indices $\{k,l\}$ \citep{Batchelor_2000}, we obtain

\begin{equation}\label{eq:viscosity_tensor}
	\eta_{ijkl} = \frac{3\mathcal{A}}{32}\left(\epsilon_{ijk}\, n_l + \epsilon_{ijl}\, n_k\right).
\end{equation}

Equation~\eqref{eq:viscosity_tensor} constitutes one of the central results of this work: it gives the emergent viscosity tensor of an active chiral film derived entirely from the microscopic orientational dynamics, without use of any phenomenology. A key structural feature of this tensor is that it is antisymmetric under the exchange  $i \leftrightarrow j$ owing to the Levi-Civita tensor associated with an antisymmetric stress. Unsurprisingly, the viscosity tensor contains signatures of the already established odd response; for the specific imposed shear flow in the $xz$ plane above the no-slip surface at $z=0$, this is made explicit by computing the sole Hall coefficient probed by this flow,
\begin{equation}\label{eq:etao}
    \eta^o_{2313} = -\eta^o_{1323} \equiv \eta^o =  \frac{3 \mathcal{A}}{32},
\end{equation}
which reflects the lack of major symmetry, a characteristic of odd viscosity, and reaffirms that the transverse traction serves as a simple scalar probe of odd rheological signatures. A carpet of orientable spinning particles thus behaves, at the macroscale, as an odd fluid.

It is interesting to interpret the above result from a complementary perspective that connects to suspension rheology of passive rods \citep{leal1971effect, hinch1972effect}. Classical formulations of rod rheology are naturally posed on the real projective space $\mathbb{RP}^2$, since Jeffery's equation is invariant under $\bm{p} \to -\bm{p}$. Anchoring to the wall breaks this nematic symmetry by restricting orientations to the upper hemisphere $\mathcal{H}$; nevertheless, the $\mathbb{RP}^2$ construction remains a useful reference problem, in which the viscosity tensor admits a compact geometric-moment representation, as we now illustrate. To this end, we replace the no-flux condition by a reappearing boundary condition, identifying antipodal points $\bp \sim  -\bp$ at the equator. Geometrically, a particle reaching the equator at an azimuthal angle $\phi$ reappears at the antipodal point $\phi+\pi$; this identification is imposed through
\begin{equation}
\psi\!\left(\theta=\frac{\pi}{2},\phi\right) = \psi\!\left(\theta=\frac{\pi}{2},\phi+\pi\right).
\end{equation}
With this boundary condition, we can now perturbatively solve for the $O(Pe)$ problem to obtain
\begin{equation}\label{eq:psi1_RP2}
\psi^{(1)} = \frac{1}{4\pi} p_k p_l E_{kl}.
\end{equation}
It is interesting to note that $\psi(\bp) = \psi(-\bp)$ since $p_k p_l E_{kl}$ is even in $\bp$, consistent with the nematic symmetry imposed by the geometric construction.  Substituting this perturbative solution in \eqref{eq:active_stress} and identifying the result with the linear constitutive relation $\Sigma_{ij} = \eta_{ijkl} E_{kl}$, the viscosity tensor can be identified as
\begin{equation}\label{eq:etarp2}
\eta_{ijkl}  = \frac{\mathcal{A}}{8\pi} \epsilon_{ijq} \int_{\mathcal{H}}  p_q p_k p_l \ \md \bp.
\end{equation}
Equation~\eqref{eq:etarp2} provides a closed-form expression for the viscosity tensor in terms of geometric moments over the hemispherical domain and further reveals its $\{k,l\}$-symmetric structure, inherited from the dyad $p_k p_l$. As in the previous calculation, this tensor contains an odd-viscous component; for the imposed shear in the $xz$ plane, one again obtains $\eta^o_{2313}=-\eta^o_{1323}\equiv\eta^o_{\mathbb{RP}^2}$. Evaluating the integral in \eqref{eq:etarp2} explicitly yields $\eta^o_{\mathbb{RP}^2} = \mathcal{A}/32$, a factor of three smaller than the exact value obtained from the no-flux calculation — a discrepancy that directly reflects the approximate treatment of the equatorial boundary in this geometric construction. 
Interestingly, the integral in \eqref{eq:etarp2} involving a third-order tensorial moment of $\bp$ vanishes identically over $\mathbb{S}^2$ because $p_qp_kp_l$ is odd under $\bp\to-\bp$. Thus, our result further highlights the role of the restricted orientational domain: in bulk suspensions, the symmetry $\bp\sim-\bp$ eliminates all odd geometric moments of $\bp$, whereas in the anchored chiral film, hemispherical orientational constraint leads to a finite polarization giving rise to signatures of Hall viscosity.

\subsection{Beyond linear response}

Having characterized the linear-response regime and the emergent odd-viscous coefficient $\eta^o$, we now ask how the response of the chiral film evolves beyond the perturbative regime at large $Pe$, when shear-induced reorientation is no longer a small correction to rotational diffusion. To explore this regime, we numerically solve the full orientational dynamics on $\mathcal{H}$ from Eqs.~\eqref{eq:normalsation_psi}--\eqref{eq:pdot_Jeffery}, together with the no-flux boundary condition at the equator. To this end, we use a spectral collocation method to represent $\psi(\theta,\phi,t)$, where a Fourier representation is used along the periodic azimuthal angle $\phi$ and a Chebyshev basis is used for the bounded polar angle $\theta$. All simulations are initialized from an isotropic state $\psi = 1/2 \pi$; the unsteady conservation \eqref{eq:psi_conservation} is then evolved using a fourth-order accurate Runge--Kutta scheme until a steady state is reached. For each value of $Pe$, we compute the steady distribution, from which the shear-induced polarity along the flow direction $P_x$ and the corresponding transverse traction $t_y =\mathcal{A} P_x/2$ are evaluated.

\begin{figure*}[hbt]
    \centering
    \includegraphics[width=1\linewidth]{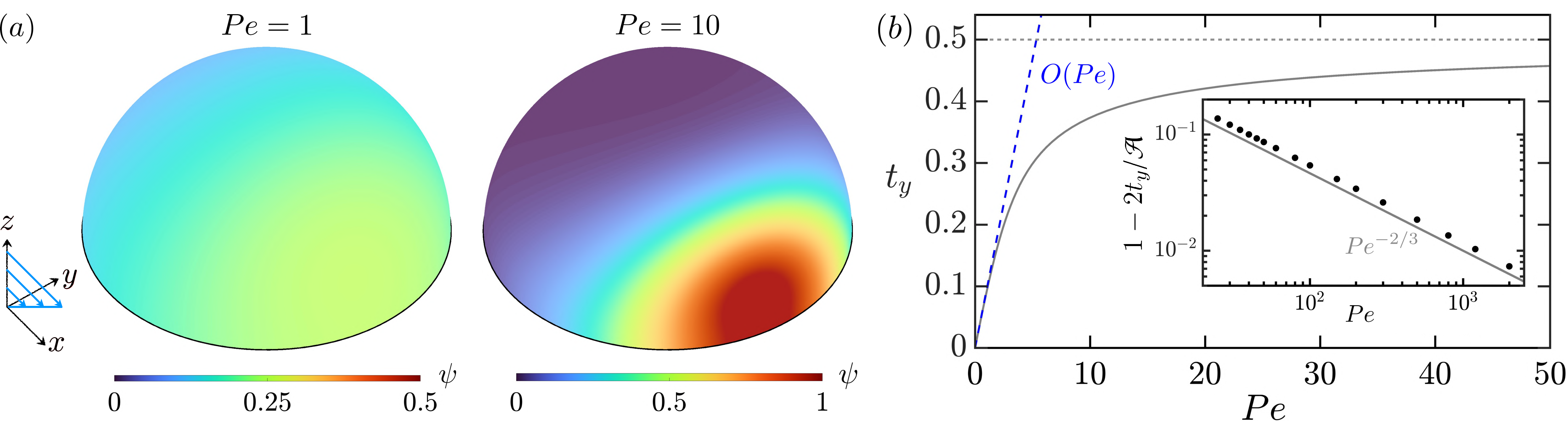}
    \caption{Nonlinear response of the active chiral film for $\mathcal{A} = 1$. (a) Steady-state orientation distributions $\psi(\bp)$ at $Pe=1$ and $10$. The blue arrows indicate the imposed simple shear, $\bu= Pe\, z\hat{\bx}$, in dimensionless variables; increasing $Pe$ concentrates particle orientations along the flow direction. (b) Emergent transverse traction $t_y$ versus $Pe$. The numerical results agree with the $O(Pe)$ prediction at weak shear, then become sublinear and approach saturation at larger $Pe$. The grey dashed line indicates the asymptotic plateau value, $t_y = \mathcal{A}/2$, in the $Pe \rightarrow \infty$ limit. In the inset, $1-2t_y/\mathcal{A}$ is plotted against $Pe$ to show the asymptotic scaling at large but finite $Pe$. Our prediction of $Pe^{-2/3}$ from a boundary layer analysis (solid line) is in excellent agreement with the numerical results (symbols). }
    \label{fig:fig3_nonlin_sim}
\end{figure*}

The steady-state distributions for two representative values of $Pe$ are shown in figure~\ref{fig:fig3_nonlin_sim}(a). For moderate shear, the distribution is only weakly biased relative to the isotropic state, consistent with the perturbative calculation. As $Pe$ increases, the distribution becomes increasingly concentrated along the shear direction, with particles accumulating in a thin orientational boundary layer near the equator $\theta = \pi/2$. This boundary-layer accumulation reflects the fact that Jeffery rotation drives particles toward wall-parallel alignment, while the no-flux condition prevents probability from crossing into the lower hemisphere. As before, we use the transverse traction $t_y$ as our diagnostic of the odd response, now tracked across the full range of $Pe$ as shown in figure~\ref{fig:fig3_nonlin_sim}(b). At low $Pe$, the numerical results are in excellent agreement with the linear-theory prediction given by \eqref{eq:ty_linear}, validating the perturbative analysis. However, as $Pe$ increases, $t_y$ departs from this linear growth and saturates to a constant value. This saturation admits a quantitative asymptotic description. As $Pe \to \infty$, the orientation distribution collapses onto the flow direction as $\psi \to \delta[\bm p - \hat{\bm x}]$, corresponding to complete alignment at $\theta = \pi/2$, $\phi = 0$. The transverse traction then attains its upper bound, $t_y \to \mathcal{A}/2$, indicated by the grey dashed line in figure~\ref{fig:fig3_nonlin_sim}$(b)$. For finite but large $Pe$, a thin boundary layer of angular thickness $\epsilon_\text{BL}$ develops around this fully aligned state. Inside this boundary layer near the wall, the advective and diffusive fluxes balance, which yields $\epsilon_\text{BL} \sim Pe^{-1/3}$, the same as that obtained in the classical asymptotic analyses of shear-aligned suspensions~\citep{hinch1972effect}. Since $p_x^2 = 1 - p_y^2 - p_z^2$ and $p_y, p_z \sim O(\epsilon_\text{BL})$, one finds $1 - P_x \sim \langle p_y^2 + p_z^2 \rangle \sim Pe^{-2/3}$ --- a scaling confirmed by our numerical results for $t_y$ (see inset of figure~\ref{fig:fig3_nonlin_sim}(b)).

The above saturation marks the crossover out of the linear-response regime: the coefficient $\eta^o$, defined in the low-${Pe}$ limit using \eqref{eq:etao}, no longer provides a meaningful characterization of the response at finite shear. At large ${Pe}$, the appropriate observable is instead the nonlinear transverse traction itself, which approaches a finite plateau. Thus, beyond the regime of constant odd viscosity, the chiral film retains a nonlinear Hall-like mechanical response: imposed shear continues to generate a finite transverse particle traction, but its magnitude is controlled by nonlinear orientational accumulation rather than by a linear constitutive coefficient.

\section{Concluding remarks}\label{sec:conclusion}

In this work, we have developed a first-principles, bottom-up derivation of emergent odd viscosity in an active chiral film, starting from the orientational dynamics of individual torque-exerting elongated particles, anchored to a no-slip surface. Unlike phenomenological descriptions of odd fluids \citep{ganeshan2017odd,soni2019odd}, which postulate an odd-viscous constitutive law, our microscopic model contains no such response \textit{a priori}. An isotropic particle distribution inside the active film of torque monopoles produces only a spatially uniform, traction-free antisymmetric stress. A finite odd-viscous response emerges only after coarse-graining the shear-driven reorientation of the particles and accounting for the geometric constraint imposed by the wall. Beyond linear response, our results highlight that the transverse traction saturates at large $Pe$, signaling a breakdown of $\eta^o$ as a linear-response coefficient while the carpet retains a nonlinear, Hall-like mechanical response. Confining the chiral medium to a wall-bounded layer further renders this response directly accessible; in confined geometries the odd stress is transmitted to the bounding walls as a transverse traction, while in open geometries it enters as a stress jump above the active film and drives potentially measurable interfacial and bulk flow, as discussed below. 

We note that our derivation shares a common structure with the microscopic coarse-graining of \citep{markovich2024nonreciprocity}; in both approaches, the odd viscosity is set by the ratio of an active torque density to a rotational relaxation rate and requires a nonequilibrium steady state sustained by activity. The two routes differ, however, in the degree of freedom through which the flow enters. In the bulk chiral fluids considered by \citep{markovich2024nonreciprocity}, the injected torque density, which sets the magnitude and axis of the steady spin density, is insensitive to the flow. In our anchored film, by contrast, the active torque density is itself the flow-responsive quantity: shear reorients the particles and generates the flow-aligned polarization from which the odd response follows. A further distinction lies in the composition of the two systems: in our film, the torque-exerting particles are embedded within an isotropic Newtonian solvent, and the odd response arises as a particle extra stress in the Stokesian fluid; whereas in the bulk chiral fluid, the constituents themselves carry the hydrodynamic momentum. We finally highlight that our film possesses a continuous rotational symmetry about the wall normal ($\uvc{z}$ in the present case) as assumed in the classification of parity-violating viscosities of \citep{khain2022stokes}. Antisymmetric stresses of the kind obtained here are admitted within that framework; the present calculation delineates a microscopic origin for such a response. \textit{In toto}, our work reveals how odd viscosity in active chiral films emerges self-consistently as a coarse-grained rheological signature of the underlying fluid--structure interaction.

Bacterial carpets \citep{Darnton2004BacterialCarpets, dauparas2018helical} provide a natural and experimentally accessible test bed for these predictions, with the transverse traction $t_y$ serving as a direct diagnostic of the odd response. Using our estimate of $\mathcal{A}$ for such carpets, and assuming the slender flagellum to behave like a rod-like particle, we predict $\eta^o\sim(5$--$10)\times\mu$, a signal large enough to be measured in future experiments. Three remarks delimit this estimate. First, it is a linear-response result, valid for shear rates $\dot\gamma \lesssim d_r$; beyond this regime the transverse traction saturates, as quantified in \S3.2. Second, the experimentally reported carpet densities can lie beyond the strictly dilute regime of our kinetic theory; however, in analogy to the rheology of passive rod suspensions, we expect our predictions to persist up to corrections of order unity \citep{batchelor1971stress}. Finally, we note that $\eta^o$ is a material property of the active layer itself, supported on $0 < z \lesssim \ell$, and the Hall signature identified here is correspondingly a thin-film effect. This quasi-two-dimensional setting is, however, suitable for rheological measurements: a colloidal probe translated within the active layer, parallel to the carpet, directly samples the odd environment and is expected to experience a transverse force. Possibilities of such particle-resolved microrheology closely parallel recent measurements in two-dimensional frictional chiral baths \citep{goerlich2026particle}, which exploit transverse displacements of tracer probes as a measure of Hall transport.
A tethered, spinning helical flagellum also generates an active force-monopolar contribution due to its geometric chirality and an associated reaction traction on the surface. However, because the shear-induced polarity is directed along $\hat{\bm x}$, this contribution modifies the longitudinal traction $t_x$ at leading order, while leaving the transverse signal $t_y$ --- our scalar probe for odd response --- unaffected. 

More broadly, while the present analysis was restricted to the dilute limit under an externally imposed flow, the active chiral film itself constitutes a rich and largely unexplored active matter system \citep{furthauer2012active}.  In a fully coupled problem, the odd stress can enter through a traction-matching condition at the interface between the active film and the passive bulk fluid above it, where the antisymmetric stress derived here appears as a stress jump and directly sources flows in the bulk. Alternatively, for a free-slip surface, the wall no longer enforces $\bm u=0$, and the odd stress can instead appear directly as a transverse traction boundary condition on the fluid itself. Such formulations should also retain the force-monopole contribution to the flow: because, in contrast to unbounded bulk fluids, force and torque monopoles share the same far-field hydrodynamic signatures near a wall, and neither hydrodynamic singularity can be neglected in computing the flow the carpet itself generates. Accounting for this disturbance flow, which in turn alters the particle orientations considered here, raises a host of new hydrodynamic questions, including spontaneous flow generation, orientational instabilities, active pumping, as well as possible additional odd responses arising at higher order in concentration: directions we wish to explore in future work.

\vspace*{1mm}

\section*{Acknowledgements}
B.C. acknowledges the support of the Department of Atomic Energy, Government of India, under project no. RTI4019, ANRF ECRG grant 002268, and support from the ICTS CALIBRE grant. S.C. and B.C. thank Sriram Ramaswamy for helpful discussions.

\bibliographystyle{ieeetr}
\bibliography{jfm.bib}

\end{document}